\documentclass[preprint,pteplogo]{ptephy_v2}
\usepackage[a4paper,margin=25mm]{geometry}
\usepackage{graphicx}
\usepackage{amsmath,amssymb,bm,mathtools}
\usepackage{hyperref}
\usepackage{subcaption}
\newcommand{\dd}{\mathrm d}

\newcommand{\beq }{\begin{equation} }
\newcommand{\eeq }{\end{equation} }

\begin{document}

\title{Phase Encoding of Genuine Three-Body Interactions \\in a Relativistic Dirac System 
\\in $1+1$ Dimensions}
\author{Tadashi Yoshikawa\\Nagoya Aoi University, Nagoya, Japan}

\begin{abstract}
We show how genuine three-body phase information can enter the invariant mass
of a relativistic three-particle Dirac system in $(1+1)$ dimensions.
As a solvable reference system, we consider the Sakamoto--Munakata--Ino model
with pairwise contact interactions
$g_{ij}(1-\alpha_i\alpha_j)\delta(x_i-x_j)$.
These singular interactions can be transferred into sector-dependent phases
and matching conditions by a discontinuous unitary transformation.
Although the explicit contact terms are thereby removed,
the nonzero constituent-mass operator is rotated and retains nontrivial spectral information.

We introduce a genuine three-body holonomy generated by
$Q_3=\alpha_1\alpha_2\alpha_3$.
The kinetic and pair-interaction parts commute with $Q_3$,
while the constituent-mass operator anticommutes with it.
Consequently, the massless system separates into the $Q_3=\pm1$ sectors,
which acquire opposite holonomy phases $e^{\pm i\theta_3}$,
whereas nonzero constituent masses mix the two sectors.
This phase-sector-mixing mechanism makes the relative three-body phase
dynamically accessible to the bound-state spectrum and establishes an
operator-level mechanism through which the three-body holonomy generates
a $\theta_3$ dependence of the physical three-body invariant mass.

We further emphasize that the topological three-body holonomy
is not automatically equivalent to a bare triple-contact potential;
such an equivalence requires a regulated self-adjoint realization
and a compatible interaction-dependent boost satisfying the Poincaré algebra.
The resulting framework therefore connects genuine three-body phase information
to the mass spectrum of a relativistic composite system while clearly separating
the controlled holonomy construction from the unresolved short-distance
triple-contact realization.
\end{abstract}

\maketitle

\section{Introduction}
Relativistic few-body systems provide a useful theoretical setting for investigating nonperturbative structures that may 
arise in strongly interacting composite systems.  One broad physical motivation comes from hadron physics, 
where conventional baryons carry three-valence-quark quantum numbers and therefore motivate the study of 
relativistic three-body bound systems.  Constituent-quark descriptions have long treated baryons 
as relativistic or relativized three-quark systems \cite{CapstickIsgur,RichardReview,Valcarce}.  
Genuine three-body confinement terms have also been investigated as possible 
contributions to hadron spectra \cite{PepinStancu,PappStancu}, and recent work has renewed this question 
by comparing meson and baryon masses within common constituent Hamiltonians \cite{Baek2026}.

Low-dimensional few-body models should not be regarded as direct descriptions of QCD.  
They can nevertheless isolate mechanisms that are obscured in the underlying field theory.  
Exactly solvable models are valuable because relativistic kinematics, singular interactions, phase structure, 
and bound-state formation can be studied analytically and nonperturbatively.

A particularly instructive class consists of Dirac particles with contact interactions in $(1+1)$ dimensions.  
Exact bound-state solutions have been constructed for two-, three-, and higher-body systems, and relativistic covariance 
has been addressed by constructing an interaction-dependent boost generator that closes 
the Poincar\'e algebra \cite{Glockle1987,Munakata1988,Munakata1990,Ino1990,Ino1992,Sakamoto1993,SakamotoNakanoYoshikawa1993}.  
We refer to this family as the Sakamoto--Munakata--Ino class. In particular, Sakamoto constructed 
an exact $N$-body bound-state solution for Dirac particles with delta-function interactions and showed that, 
when all constituent masses vanish, the interacting model becomes unitarily equivalent to the free theory \cite{Sakamoto1993}.

For three particles, the pairwise interaction is
\begin{equation}
 V_2=\sum_{i<j}g_{ij}(1-\alpha_i\alpha_j)\delta(x_i-x_j),
 \label{eq:V2}
\end{equation}
where $\alpha_i$ acts on the spinor factor of particle $i$.  Its singular part may be encoded 
in a discontinuous unitary transformation,
\begin{equation}
 \Psi=U_2\Phi,\qquad U_2=e^{i\Theta_2}.
\end{equation}
In the appropriate self-adjoint realization, derivatives of $\Theta_2$ reproduce the contact terms.  
The central structural relation is
\begin{equation}
 U_2^{-1}(H_0+V_2)U_2=H_0+\Delta H_m.
 \label{eq:central}
\end{equation}
Thus an interaction can disappear as an explicit distribution and still survive through rotated mass matrices and matching data.  
When all constituent masses vanish, $\Delta H_m$ can vanish and the system becomes unitarily equivalent to the free theory. 
 For nonzero masses, $\Delta H_m$ is generally nonzero and the phase information can affect the invariant bound-state mass.

The three pair-coincidence sets intersect at $x_1=x_2=x_3$.  This suggests adding
\begin{equation}
 V_3=g_3Q_3\delta(x_1-x_2)\delta(x_2-x_3),\qquad
 Q_3=\alpha_1\alpha_2\alpha_3.
 \label{eq:V3}
\end{equation}
At the same time, removal of the triple-coincidence line from the full configuration space leaves a puncture 
in the two-dimensional relative space.  Its fundamental group is $\mathbb Z$, so a wave function may acquire 
an independent holonomy $e^{i\theta_3}$ around the puncture \cite{HarshmanKnapp,Ohya2024}.  The purpose of this paper 
is to determine precisely what can be inferred from this phase and to separate it from claims that require 
a regularized operator realization of Eq.~\eqref{eq:V3}.

The central observation is algebraic.  The kinetic and Sakamoto pair-interaction operators preserve 
the eigenspaces of $Q_3$, whereas the constituent-mass operator interchanges them.  A matrix-valued three-body holonomy 
assigns opposite phases to the two sectors.  Consequently, the constituent masses provide the channel through 
which the relative phase between $Q_3=+1$ and $Q_3=-1$ can affect a composite eigenvalue.  
This phase--sector--mixing mechanism, rather than the determinant condition by itself, is the organizing principle of the present work.

The paper is organized as follows. Section II defines the relativistic
three-body system and states the Poincar\'e-algebra constraints.
Section III reviews the phase transformation of the pair contacts and
the residual rotation of the mass operator. Section IV distinguishes a
triple-contact potential from a topological three-body holonomy.
Section V compares the massless and massive limits and gives a
phenomenological interpretation. Section VI develops the $Q_3$-sector
decomposition and mass-induced mixing. Section VII constructs the
six-sector loop condition and explains its status as a general, not yet
evaluated, spectral equation. Section VIII discusses the resulting mass
mechanism, discrete symmetries, and possible flavor extensions.
The conclusion is followed by two appendices: an alternative three-body sector projector and a brief $N$-body extension.

\section{Three-body Dirac model and Poincar\'e algebra}
The free Hamiltonian and total momentum of the three-body Dirac model are 
\begin{equation}
 H_0=\sum_{i=1}^3(\alpha_i p_i+\beta_i m_i),\qquad
 P=\sum_{i=1}^3p_i,\qquad p_i=-i\partial_{x_i}.
 \label{eq:H0}
\end{equation}
Here we use units $\hbar=c=1$ and the Dirac matrices are 
\begin{align}
 \alpha_i = \begin{pmatrix} 0&1\\1&0 \end{pmatrix}, \qquad 
    \beta_i=\begin{pmatrix} 1&0\\0&-1 \end{pmatrix}
\end{align}
which satisfy
\begin{equation}
 \alpha_i^2=\beta_i^2=1,\qquad \{\alpha_i,\beta_i\}=0,
\end{equation}
with matrices belonging to different particles commuting.  
The formal interacting Hamiltonian is 
\begin{equation}
  H=H_0+V_2+V_3.
\end{equation}
where $V_2$ and $V_3$ are shown in Eqs.~\eqref{eq:V2} and \eqref{eq:V3}.

At $t=0$, the $(1+1)$-dimensional Poincar\'e algebra is
\begin{equation}
 [H,P]=0,\qquad [K,P]=iH,\qquad [K,H]=iP.
 \label{eq:palg}
\end{equation}
The free boost may be written in symmetrized form
\begin{equation}
 K_0=\frac12\sum_i\{x_i,h_i\},\qquad h_i=\alpha_ip_i+\beta_im_i,
 \label{eq:K0}
\end{equation}
up to the overall sign convention for boosts.  It satisfies Eq.~\eqref{eq:palg} with $H_0$.

Write the interacting boost as
\begin{equation}
 K=K_0+Z,\qquad Z=Z_2+Z_3.
\end{equation}
Closure of the algebra is equivalent to
\begin{align}
 [V,P]&=0, \label{eq:trans}\\
 [Z,P]&=iV, \label{eq:boost1}\\
 [K_0,V]+[Z,H_0]+[Z,V]&=0, \label{eq:boost2}
\end{align}
where $V=V_2+V_3$.  Equation~\eqref{eq:trans} follows for Eqs.~\eqref{eq:V2} and \eqref{eq:V3} 
because they depend only on coordinate differences.  Equations~\eqref{eq:boost1}--\eqref{eq:boost2}, 
however, are additional dynamical constraints.  For the established pairwise model, $Z_2$ is 
the interaction-dependent boost correction constructed in the exact-solution literature \cite{Munakata1988,Munakata1990,Ino1990,Ino1992,Sakamoto1993,SakamotoNakanoYoshikawa1993}.  
Once $V_3$ is added, covariance requires an operator $Z_3$ satisfying the terms linear 
and nonlinear in $V_3$ in Eqs.~\eqref{eq:boost1}--\eqref{eq:boost2}.  Translation invariance alone 
is not a proof of Poincar\'e covariance.

This observation supplies a clean consistency test.  A regularized three-body interaction $V_3^{(\epsilon)}$ defines 
a relativistic model only if one can construct $Z_3^{(\epsilon)}$ on a common invariant domain and obtain 
a regulator-independent limit of the commutators.  Alternatively, one may formulate the interaction directly 
as covariant matching data and prove that the boosted matching conditions have the same form.  
We shall use the second, holonomy-based description below and keep the bare potential \eqref{eq:V3} 
as a formal representative until this test is completed.

\section{Pairwise phase transformation and the mass sector}
\begin{figure*}[t]
    \centering
    \begin{subfigure}[t]{0.48\textwidth}
        \centering
        \includegraphics[width=\linewidth]{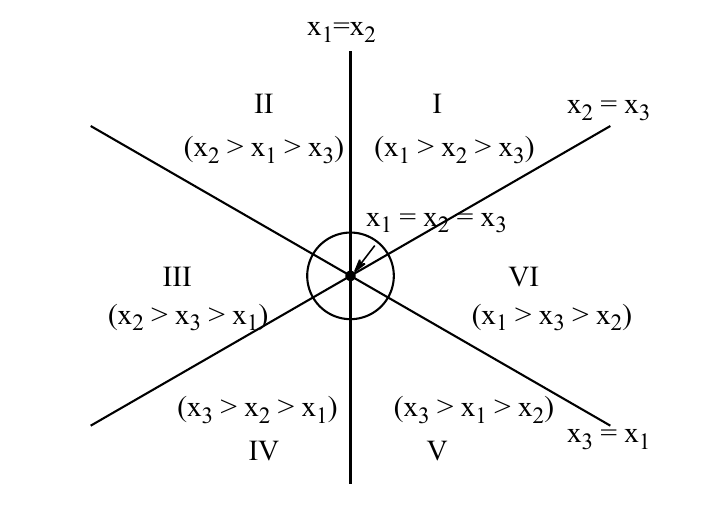}
        \caption{
        Six ordering sectors corresponding to the possible
        orderings of the three particle coordinates.
        }
        \label{fig:sector-ordering}
    \end{subfigure}
    \hfill
    \begin{subfigure}[t]{0.48\textwidth}
        \centering
        \includegraphics[width=\linewidth]{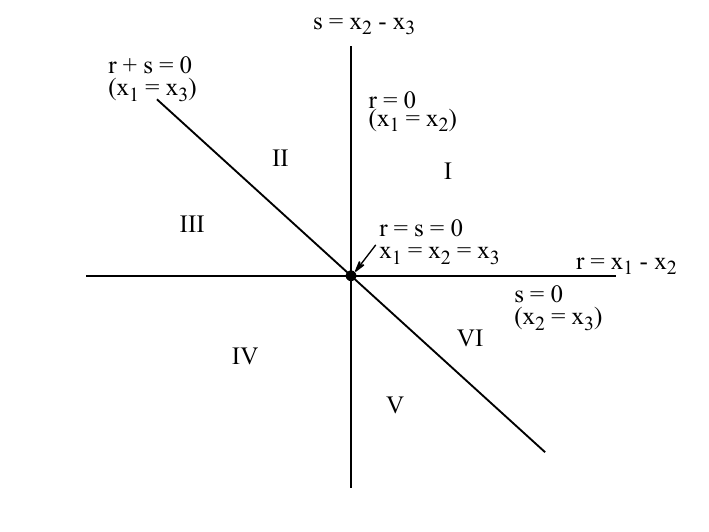}
        \caption{
        Relative configuration space in the coordinates
        $r=x_1-x_2$ and $s=x_2-x_3$.  The three coincidence
        lines $x_1=x_2$, $x_2=x_3$, and $x_1=x_3$ divide the
        plane into the six ordering sectors.
        }
        \label{fig:relative-space}
    \end{subfigure}
    \caption{
    Configuration-space structure of the three-particle system.
    (a) The six sectors corresponding to the possible particle
    orderings.  (b) The same sector structure represented in the
    relative-coordinate plane $(r,s)$.  The three pair-coincidence
    lines intersect at $r=s=0$, corresponding to the triple-coincidence
    point $x_1=x_2=x_3$.  This point is the distinguished point around
    which the three-body holonomy is defined.
    }
    \label{fig:configuration-space}
\end{figure*}

The phase-transformation viewpoint used below is closely connected with Sakamoto's exact $N$-body construction \cite{Sakamoto1993}.  
The relation between two forms of the $N$-body bound-state solutions, including the role of different treatments of the delta-function interaction, 
was subsequently clarified by Sakamoto, Nakano, and Yoshikawa \cite{SakamotoNakanoYoshikawa1993}.
Let the pair-coincidence lines divide the relative configuration plane into six ordering sectors,
 as illustrated in Fig.~\ref{fig:configuration-space}.
 Away from those lines, $V_2=0$ and the wave function obeys the free Dirac equation.  Across $x_i=x_j$, 
 integration of the equation through a thin transverse strip produces a unitary matching matrix $S_{ij}(g_{ij})$.  
 Equivalently, one may choose a piecewise constant phase $U_2$ whose jump is $S_{ij}$.
The pairwise contact interactions connect the six ordering sectors through two-body matching matrices
 $S_{ij}$. If $A_{\sigma}$ denotes the coefficient vector in a given sector, 
 crossing the coincidence line $x_i=x_j$ gives schematically 
 \begin{equation} 
  A_{\sigma'} = S_{ij}(E,g)\,A_{\sigma}, 
\end{equation} 
where $\sigma'$ differs from $\sigma$ by the interchange of particles $i$ and $j$. 
For a purely factorized three-body problem, the consistency of different sequences of pairwise exchanges 
is expressed by the Yang--Baxter relation\cite{Yang}, schematically, 
\begin{equation}
   S_{12} S_{13} S_{23} = S_{23} S_{13} S_{12}. 
\end{equation} 
In the present work, this relation is mentioned only as the standard consistency condition for factorized pairwise sector matching; 
our main interest is the additional genuine three-body phase and its effect on the mass condition.

For any differentiable portion of $U_2$,
\begin{equation}
 U_2^{-1}H_0U_2
 =H_0+\sum_i\alpha_i A_i
 +\sum_i m_i\left(U_2^{-1}\beta_iU_2-\beta_i\right),
 \qquad A_i=-iU_2^{-1}\partial_iU_2.
 \label{eq:gauge-transform}
\end{equation}
The singular part $\sum_i\alpha_iA_i$ cancels $V_2$ by construction.  The remaining term is
\begin{equation}
 \Delta H_m=\sum_i m_i\left(U_2^{-1}\beta_iU_2-\beta_i\right),
 \label{eq:DeltaHm}
\end{equation}
which establishes Eq.~\eqref{eq:central}.  This is the precise reason that phase absorption 
does not generically remove spectral effects.

For illustration, if a component of the transformation is generated by a Hermitian 
involution $Q$ with $Q^2=1$ and $\{Q,\beta_i\}=0$, then
\begin{equation}
 e^{-i\vartheta Q}\beta_i e^{i\vartheta Q}
 =\beta_i\cos(2\vartheta)+i\beta_iQ\sin(2\vartheta).
 \label{eq:massrotation}
\end{equation}
For $Q_3=\alpha_1\alpha_2\alpha_3$, one has $Q_3^2=1$ and $\{Q_3,\beta_i\}=0$ for each $i$.  
Consequently, any legitimate $Q_3$-generated three-body phase rotates every nonzero mass term.  
This algebraic statement is exact; determining the allowed $\vartheta$ requires the triple-coincidence domain 
or matching condition.


\section{Three-body interaction, holonomy, and mass modification}

We now introduce the genuine three-body structure considered in this work.
In contrast to the pairwise contact interactions $V_{ij}$, which act on the
two-particle coincidence lines $x_i=x_j$, a genuine three-body contribution
is associated with the collective internal structure of the three-particle
system and cannot, in general, be reduced to a sum of pairwise terms.

It is important, however, to distinguish an explicit three-body contact
potential from a three-body holonomy.  In the present work, we characterize
the latter by the unitary operator
\begin{equation}
    U_3(\theta_3)=\exp(i\theta_3 Q_3),
\end{equation}
where $\theta_3$ denotes a three-body phase and $Q_3$ acts simultaneously
on the internal degrees of freedom of the three particles.

The central observation is that a local representation of such a phase
acts nontrivially on the massive sector.  Since $Q_3$ commutes with the
kinetic matrices $\alpha_i$ but anticommutes with the mass matrices, a
local gauge representation of the holonomy rotates the matrix form of the
mass operator.  This local rotation alone is a unitary change of
representation and therefore does not change the spectrum.  A physical
$\theta_3$ dependence can arise only when the phase is part of the global
three-body domain or matching condition around the puncture.  In that case
the bound-state mass may be written schematically as
\begin{equation}
    M_3=M_3(\boldsymbol g,\theta_3;m_1,m_2,m_3),
\end{equation}
with the actual dependence determined by the completed matching problem.

We next clarify the geometrical meaning of $\theta_3$ and its relation to
triple coincidence.

Introduce the relative coordinates
\begin{equation}
    r=x_1-x_2,\qquad s=x_2-x_3,
\end{equation}
together with a center coordinate $X$.  Triple coincidence corresponds to
the single point $(r,s)=(0,0)$ in the relative configuration space.
Removing this point gives
\begin{equation}
    \pi_1\!\left(\mathbb R^2\setminus\{0\}\right)=\mathbb Z.
\end{equation}
Therefore, a one-dimensional unitary representation of the fundamental
group may be characterized by
\begin{equation}
    \Phi(\rho,\varphi+2\pi)
    =
    e^{i\theta_3}\Phi(\rho,\varphi),
    \qquad
    \theta_3\equiv\theta_3+2\pi .
    \label{eq:twist}
\end{equation}

Equivalently, the same holonomy may be represented locally by a flat
Abelian connection outside the origin,
\begin{equation}
    A^{(3)}
    =
    \frac{\theta_3}{2\pi}\,\dd\varphi,
    \qquad
    \oint A^{(3)}=\theta_3,
\end{equation}
whose singular curvature is concentrated at the triple-coincidence point,
\begin{equation}
    \dd A^{(3)}
    =
    \theta_3\,
    \delta^{(2)}(r,s)\,
    \dd r\wedge\dd s .
    \label{eq:AB}
\end{equation}
This construction is the configuration-space analogue of 
the Aharonov--Bohm mechanism, in which a locally flat connection 
can produce a physically nontrivial holonomy around an excluded 
region \cite{AB}. In the present three-body problem, the excluded 
triple-coincidence point plays the role of the puncture, 
while the associated one-dimensional contact topology has been discussed 
in Refs. \cite{HarshmanKnapp,Ohya2024}.

At this point, a distinction is essential.  A unitary transformation of a
first-order Dirac kinetic operator generates a connection term of the form
$\sum_i\alpha_i A_i$, rather than directly producing a multiplicative
scalar interaction localized at the triple-coincidence point.  The
point-supported object in Eq.~\eqref{eq:AB} is the curvature of the
singular connection, whereas the Hamiltonian couples to the connection
itself.  Hence,
\begin{equation}
    g_3 Q_3\delta(r)\delta(s)
    \quad\not\equiv\quad
    \text{three-body holonomy}
    \label{eq:nonequiv}
\end{equation}
without specifying a regularization and demonstrating that the two
descriptions define the same self-adjoint extension.

Thus, the phase $\theta_3$ is well defined through the boundary condition
\eqref{eq:twist}, whereas a universal regulator-independent relation
$\theta_3=F(g_3)$ should not be assumed.

For the matrix-valued case relevant to the present model, we impose
\begin{equation}
    \Phi(\varphi+2\pi)
    =
    e^{i\theta_3Q_3}\Phi(\varphi).
    \label{eq:matrix-twist}
\end{equation}
Because $Q_3$ commutes with the $\alpha_i$ but anticommutes with the
$\beta_i$, the kinetic structure is preserved under this transformation,
whereas the massive sector is rotated as discussed below.

\section{Massless and massive limits: phenomenological interpretation}

The role of the three-body phase becomes particularly transparent by
comparing the massless and massive limits.  The free three-particle
Dirac Hamiltonian may be written as
\begin{equation}
    H_0
    =
    \sum_{i=1}^{3}
    \left(
        -i\alpha_i\partial_i
        +m_i\beta_i
    \right).
\end{equation}
For the three-body generator considered here, we assume
\begin{equation}
    [Q_3,\alpha_i]=0,
    \qquad
    \{Q_3,\beta_i\}=0.
\end{equation}
The first relation implies that the kinetic sector is unchanged by the
three-body transformation,
\begin{equation}
    U_3^{-1}\alpha_i U_3=\alpha_i.
\end{equation}
By contrast, the mass matrices are rotated.  For $Q_3^2=1$,
\begin{equation}
    U_3^{-1}\beta_i U_3
    =
    \beta_i\cos(2\theta_3)
    +i\beta_iQ_3\sin(2\theta_3).
    \label{eq:massrotation3}
\end{equation}

\subsection{Massless limit}

In the massless limit,
\begin{equation}
    m_i\rightarrow 0,
\end{equation}
the explicit mass sector disappears from the free Hamiltonian.  Since
$Q_3$ commutes with the kinetic matrices, the transformation generated by
$Q_3$ does not modify the local kinetic operator.  The phase
$\theta_3$ may still characterize a nontrivial global boundary condition
or holonomy in configuration space, but there is no mass term on which
the rotation in Eq.~\eqref{eq:massrotation3} can act directly.

Thus, within the present mechanism, the massless system provides a useful
reference limit: the three-body phase can remain as geometrical
information, while its direct manifestation as a modification of the
mass sector disappears.

\subsection{Massive case}

For nonzero constituent masses, the situation is qualitatively different.
The transformation generated by $Q_3$ leaves the kinetic sector unchanged
but rotates the mass operator,
\begin{equation}
    \sum_i m_i\beta_i
    \longrightarrow
    \sum_i m_i
    \left[
        \beta_i\cos(2\theta_3)
        +i\beta_iQ_3\sin(2\theta_3)
    \right].
\end{equation}
When the same phase is imposed as nontrivial global matching data, the
mass-induced coupling between the $Q_3$ sectors provides a channel through
which it can enter the spectral condition, even though the local kinetic
matrices are unchanged.

Schematically, the resulting three-body bound-state condition may be
written as
\begin{equation}
    {\cal F}\bigl(M_3,g,\theta_3;m_1,m_2,m_3\bigr)=0,
\end{equation}
and hence
\begin{equation}
    M_3=M_3(g,\theta_3;m_1,m_2,m_3).
\end{equation}
The precise functional dependence is model dependent, but the structural
origin of the $\theta_3$ dependence is already apparent from
Eq.~\eqref{eq:massrotation3}.

\subsection{Phenomenological interpretation}

The comparison between the two limits suggests a simple interpretation.
The three-body phase is not itself a conventional additive mass term.
Rather, it represents collective three-body information that becomes
spectrally visible through its coupling to the massive Dirac sector.
Consequently, its effect is expected to be suppressed in the massless
limit and to become relevant when a nonzero intrinsic mass scale is
present.

This observation may be of phenomenological interest for relativistic
composite systems.  In such systems, a genuine many-body structure could
modify the physical composite mass without being represented simply as
an additional pairwise interaction.  The present $(1+1)$-dimensional
model should not be regarded as a direct model of QCD; rather, it provides
a solvable setting in which the mechanism relating a collective
three-body phase to a modification of the composite mass can be isolated.

\section{$Q_3$ sectors and mass-induced mixing}
The role of the constituent masses becomes transparent after resolving the three-particle spinor space into eigenspaces of
\begin{equation}
 Q_3=\alpha_1\alpha_2\alpha_3,
 \qquad Q_3^2=1.
\end{equation}
The corresponding projectors are
\begin{equation}
 \Pi_\pm=\frac12(1\pm Q_3),
 \qquad \mathcal H=\mathcal H_+\oplus\mathcal H_-,
 \qquad Q_3\Psi_\pm=\pm\Psi_\pm.
 \label{eq:projectors}
\end{equation}
For a two-component Dirac spinor assigned to each particle, the full spinor space has dimension eight and the two $Q_3$ sectors 
have dimension four.

Separate the Hamiltonian into the kinetic-plus-pair part and the constituent-mass part,
\begin{equation}
 H=H_\chi+H_m,
 \qquad
 H_\chi=\sum_i\alpha_ip_i+V_2,
 \qquad
 H_m=\sum_i m_i\beta_i.
 \label{eq:Hsplit}
\end{equation}
Because $Q_3$ commutes with every $\alpha_i$ and with every product $\alpha_i\alpha_j$,
\begin{equation}
 [Q_3,H_\chi]=0.
 \label{eq:Qcommute}
\end{equation}
Consequently, the kinetic and pair-interaction dynamics preserve the two eigenspaces,
\begin{equation}
 \Pi_\pm H_\chi\Pi_\mp=0.
\end{equation}

In contrast, $Q_3$ anticommutes with each $\beta_i$, and hence
\begin{equation}
\{Q_3,H_m\}=0.
 \label{eq:Qanticommute}
\end{equation}
It follows that
\begin{equation}
 \Pi_\pm H_m\Pi_\pm=0,
 \qquad
 \Pi_+H_m\Pi_-\ne0
 \quad\hbox{in general}.
 \label{eq:massoffdiag}
\end{equation}
Thus $H_m$ maps $\mathcal H_+$ into $\mathcal H_-$ and vice versa.  This statement concerns the bare constituent-mass operator; 
the unitary transformation does not change the numerical constituent masses, but it changes the matrix form in 
which they enter the transformed Hamiltonian.

\begin{figure*}[t]
    \centering

    \begin{subfigure}[t]{0.48\textwidth}
        \centering
        \includegraphics[width=\linewidth]{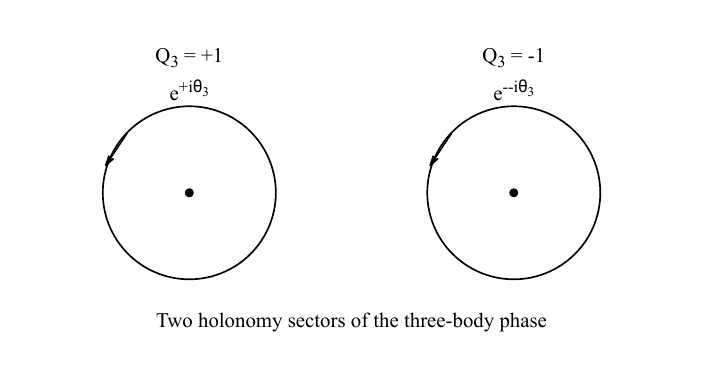}
        \caption{
        Two holonomy sectors associated with the eigenvalues
        $Q_3=\pm1$.  For the same positive orientation around the
        triple-coincidence point, the two sectors acquire the phases
        $e^{+i\theta_3}$ and $e^{-i\theta_3}$, respectively.
        }
        \label{fig:holonomy-sectors}
    \end{subfigure}
    \hfill
    \begin{subfigure}[t]{0.48\textwidth}
        \centering
        \includegraphics[width=\linewidth]{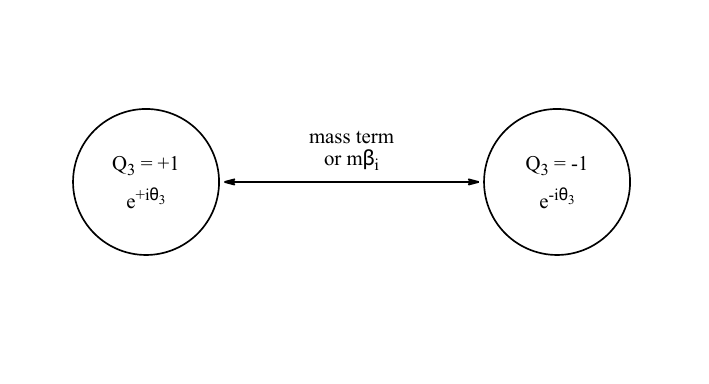}
        \caption{
        Schematic coupling between the $Q_3=\pm1$ sectors induced by
        the massive Dirac term.  The mass term connects the two
        holonomy sectors, while the kinetic sector remains unchanged.}
        \label{fig:mass-coupling}
    \end{subfigure}
    \caption{
    Role of the three-body phase in the $Q_3$ eigenbasis.
    (a) The three-body holonomy produces opposite phase factors
    $e^{\pm i\theta_3}$ in the $Q_3=\pm1$ sectors.
    (b) For nonzero mass, the Dirac mass term couples the two sectors.
    Since $[Q_3,\alpha_i]=0$ but $\{Q_3,\beta_i\}=0$, the kinetic
    structure is unchanged whereas the massive sector is rotated by
    the three-body phase.
    }
    \label{fig:holonomy-mass}
\end{figure*}

The three-body holonomy has the spectral decomposition
\begin{equation}
 W_3(\theta_3)=e^{i\theta_3Q_3}
 =e^{i\theta_3}\Pi_+ + e^{-i\theta_3}\Pi_- .
 \label{eq:holonomy-spectral}
\end{equation}
It therefore assigns opposite phases to the two sectors.  These phases would remain dynamically separated if $m_i=0$.  
For nonzero masses, Eq.~\eqref{eq:massoffdiag} couples the sectors and makes their relative phase accessible to the eigenvalue problem.
The phase-sector structure and the mass-induced coupling mechanism are
summarized schematically in Fig.~\ref{fig:holonomy-mass}.  In the
massless limit the two $Q_3$ sectors remain dynamically separated,
whereas nonzero constituent masses couple them and make their relative
holonomy phase accessible to the spectrum.

In the $Q_3$-resolved basis the Hamiltonian has the block form
\begin{equation}
 H=
 \begin{pmatrix}
  H_+ & \Delta_m\\
  \Delta_m^\dagger & H_-
 \end{pmatrix},
 \qquad
 \Delta_m=\Pi_+H_m\Pi_- .
 \label{eq:blockH}
\end{equation}
The diagonal blocks contain the kinetic, pair-interaction, and sector-dependent holonomy data, while $\Delta_m$ is entirely mass induced.  
A two-state reduction illustrates the mechanism without claiming the complete spectrum.  If $|+\rangle$ and $|-\rangle$ 
are representative states from the two sectors, then
\begin{equation}
 H_{\rm eff}(\theta_3)=
 \begin{pmatrix}
  \varepsilon_+(\theta_3)&\Delta\\
  \Delta^*&\varepsilon_-(\theta_3)
 \end{pmatrix},
 \qquad
 \Delta=\langle+|H_m|-\rangle,
 \label{eq:Heff}
\end{equation}
with eigenvalues
\begin{equation}
 E_\pm(\theta_3)=
 \frac{\varepsilon_+(\theta_3)+\varepsilon_-(\theta_3)}{2}
 \pm\sqrt{
 \left[\frac{\varepsilon_+(\theta_3)-\varepsilon_-(\theta_3)}{2}\right]^2
 +|\Delta|^2}.
 \label{eq:twolevel}
\end{equation}
Equation~\eqref{eq:twolevel} shows the possible avoided-crossing and mass-shift mechanism.  
It does not establish that the relevant diagonal splitting or $\Delta$ is nonzero for a particular bound state; 
those quantities must be computed from the full sector solutions.

\section{Six-sector matching and the invariant-mass condition}
We now explain how the $Q_3$-sector mechanism enters the global bound-state condition.  
Introduce relative coordinates
\begin{equation}
 r=x_1-x_2,
 \qquad s=x_2-x_3.
\end{equation}
The three pair-coincidence lines are $r=0$, $s=0$, and $r+s=0$.  They divide the relative plane into the six ordering sectors
\begin{equation}
 123,\quad132,\quad312,\quad321,\quad231,\quad213,
\end{equation}
where, for example, $123$ denotes $x_1<x_2<x_3$.  Their common intersection is the triple-coincidence point $(r,s)=(0,0)$.

In sector $a$, expand a normalizable solution of energy $E$ in an appropriate local basis,
\begin{equation}
 \Psi_a(r,s;E)=\sum_n(C_a)_n u_{a,n}(r,s;E),
 \label{eq:sector-expansion}
\end{equation}
and let $C_a$ denote its amplitude vector.  Crossing a pair-coincidence line gives an adjacent-sector relation
\begin{equation}
 C_{a+1}=T_{a+1,a}(E;\boldsymbol g)C_a.
 \label{eq:sector-transfer}
\end{equation}
The matrix $T_{a+1,a}$ must contain not only the local contact jump but also the change between the energy-dependent normalizable 
bases in the adjacent sectors.  For a single Sakamoto contact, we denote the corresponding local spinor jump by
\begin{equation}
 S_{ij}(g_{ij}),
 \label{eq:Sjump}
\end{equation}
with the inverse matrix used for the reverse crossing.  Its explicit form depends on the chosen self-adjoint contact prescription.  The complete $T_{a+1,a}(E;\boldsymbol g)$ contains, in addition, the change between the energy-dependent normalizable bases in adjacent sectors and is therefore more than the local jump factor alone.

Following the ordered circuit
\begin{equation}
 123\to132\to312\to321\to231\to213\to123
\end{equation}
gives
\begin{equation}
 C_{123}^{\rm after}
 =\mathcal T_{\rm loop}(E;\boldsymbol g)C_{123}^{\rm before},
 \qquad
 \mathcal T_{\rm loop}
 =T_{123,213}\cdots T_{312,132}T_{132,123}.
 \label{eq:monodromy}
\end{equation}
The three-body domain condition independently requires
\begin{equation}
 C_{123}^{\rm after}=W_3(\theta_3)C_{123}^{\rm before},
 \qquad W_3(\theta_3)=e^{i\theta_3Q_3}.
 \label{eq:holonomy-close}
\end{equation}
Compatibility of Eqs.~\eqref{eq:monodromy} and \eqref{eq:holonomy-close} gives
\begin{equation}
 [\mathcal T_{\rm loop}(E;\boldsymbol g)-W_3(\theta_3)]C_{123}=0.
 \label{eq:linear-consistency}
\end{equation}
A nonzero amplitude vector exists only if
\begin{equation}
 D(E,\theta_3;\boldsymbol g)
 :=\det[W_3^{-1}(\theta_3)\mathcal T_{\rm loop}(E;\boldsymbol g)-\mathbf1]=0.
 \label{eq:secular-matrix}
\end{equation}
Equivalently, one may write $\det[\mathcal T_{\rm loop}-e^{i\theta_3Q_3}]=0$ when both operators are represented 
on the same amplitude space.  This determinant means that local pairwise matching around the puncture 
and the global three-body holonomy admit a common nonzero state.

The $Q_3$ structure is now explicit.  If the massless loop operator commutes with $Q_3$, then
\begin{equation}
 \mathcal T_{\rm loop}^{(0)}=
 \begin{pmatrix}
  \mathcal T_+(E)&0\\0&\mathcal T_-(E)
 \end{pmatrix},
 \qquad
 W_3=
 \begin{pmatrix}
  e^{i\theta_3}\mathbf1_+&0\\0&e^{-i\theta_3}\mathbf1_-
 \end{pmatrix}.
 \label{eq:blockloop0}
\end{equation}
The consistency condition then separates into the two opposite-holonomy sectors.  With nonzero constituent masses, 
the complete loop operator can acquire off-diagonal blocks,
\begin{equation}
 \mathcal T_{\rm loop}=
 \begin{pmatrix}
  \mathcal T_{++}&\mathcal T_{+-}\\
  \mathcal T_{-+}&\mathcal T_{--}
 \end{pmatrix},
 \label{eq:blockloopm}
\end{equation}
and the off-diagonal blocks encode the mass-induced mixing identified in Eq.~\eqref{eq:massoffdiag}.

In the center-of-mass frame, a real root in the normalizable domain defines the invariant three-body mass,
\begin{equation}
  M_3(\boldsymbol g,\theta_3)=E_{\rm bound}(\boldsymbol g,\theta_3;P=0).
 \label{eq:M3def}
\end{equation}
The constituent masses $m_i$ and the composite invariant mass $M_3$ are distinct quantities.

Equation~\eqref{eq:secular-matrix} is presently a general consistency condition, not an evaluated mass formula.  
In particular, multiplying only commuting, energy-independent local jump factors can give a trivial loop product 
and cannot quantize $E$.  An explicit prediction for $M_3(\boldsymbol g,\theta_3)$ requires the normalizable sector bases, 
their energy-dependent transformations, and a specified self-adjoint triple-contact domain.

Several consequences follow immediately.  First,
\begin{equation}
 M_3(\theta_3+2\pi)=M_3(\theta_3).
\end{equation}
Second, if the completed determinant has a simple root $E_0$ at $\theta_3=0$, the implicit-function theorem yields
\begin{equation}
 \left.\frac{\partial M_3}{\partial\theta_3}\right|_0
 =-\left.\frac{\partial_{\theta_3}D}{\partial_ED}\right|_{(E_0,0)}.
 \label{eq:linearshift}
\end{equation}
If a symmetry enforces $D(E,\theta_3)=D(E,-\theta_3)$, the linear term vanishes and
\begin{equation}
 M_3(\theta_3)=E_0-\frac12
 \left.\frac{\partial_{\theta_3}^2D}{\partial_ED}\right|_{(E_0,0)}
 \theta_3^2+O(\theta_3^4).
 \label{eq:quadraticshift}
\end{equation}
This coefficient is the schematic quantity $C_2(\boldsymbol g)$.  Namely,
\begin{equation}
 C_2(\boldsymbol g)=-\frac12
 \left.\frac{\partial_{\theta_3}^2D}{\partial_ED}\right|_{(E_0,0)}.
 \label{eq:C2}
\end{equation}
Neither its sign nor its nonvanishing follows from topology alone.  Its value cannot be fixed 
before the explicit sector transfer matrices and the self-adjoint triple-contact prescription are fixed.

The role of mass is transparent.  In the massless limit the phase transformation may reduce to a change 
of sector convention, and the energy dependence on the phase can disappear.  With $m_i\ne0$, Eq.~\eqref{eq:massrotation} 
supplies a mechanism for $\partial_{\theta_3}D$ or $\partial_{\theta_3}^2D$ to be nonzero.  
Thus a three-body phase can modify the invariant mass, but its magnitude is a spectral output, not a purely topological number.

\section{Discussion}
\begin{figure}[t]
    \centering
    \includegraphics[width=0.85\linewidth]{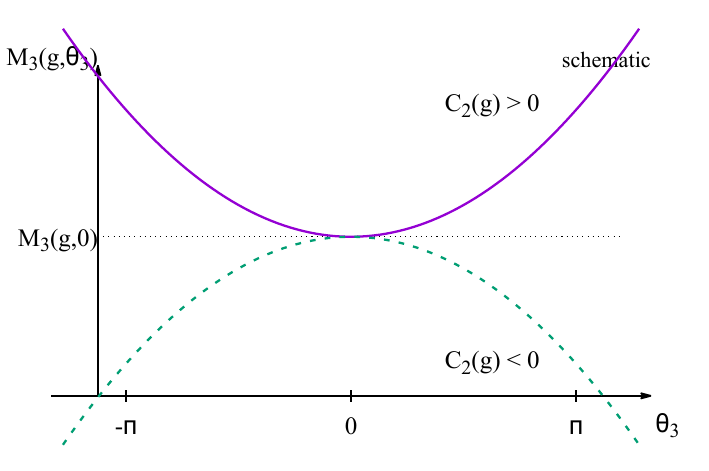}
    \caption{
    Schematic dependence of the physical three-body mass
    $M_3(g,\theta_3)$ on the three-body phase $\theta_3$.
    The curve illustrates qualitatively a possible phase dependence
    of the bound-state mass through the massive Dirac sector.
    The shape is schematic and does not represent a numerical result
    of the present model.
    }
    \label{fig:mass-phase}
\end{figure}

The central chain established by the present formulation is
\begin{equation}
 \text{three-body domain/connection}
 \longrightarrow e^{\pm i\theta_3}
 \longrightarrow Q_3\text{-sector mixing by }H_m
 \longrightarrow M_3.
\end{equation}
If the completed determinant is even under
$\theta_3\to-\theta_3$, the leading phase dependence near
$\theta_3=0$ is quadratic, as expressed in Eq.~\eqref{eq:quadraticshift}.
A schematic illustration of one possible phase dependence is shown in
Fig.~\ref{fig:mass-phase}.  The upward curvature displayed in the figure
is illustrative only; the sign of $C_2(g)$ is not determined by the
present analysis.
The essential step is not merely the appearance of a phase.  Since $H_\chi$ preserves $Q_3$, 
the holonomy gives the two sectors opposite phases.  Since $H_m$ interchanges the sectors, 
a nonzero constituent mass makes this relative phase available to the spectral problem.  
This identifies an operator mechanism for a mass effect, but not its numerical magnitude.  
Topology guarantees the admissibility and periodicity of the holonomy parameter; dynamics 
and self-adjointness determine how it is related to a regulated short-range interaction 
and whether it produces a nonzero shift for a given state.

The Poincar\'e algebra adds a separate and indispensable requirement.  
The pairwise Sakamoto--Munakata interaction belongs to a class for which an interaction-dependent boost has been constructed.  
For a new three-body term, the existence of $Z_3$ must be demonstrated, or covariance must be proved directly 
for its matching conditions.  Until then, Eq.~\eqref{eq:V3} should be described as 
a formal rest-frame contact term, while Eqs.~\eqref{eq:twist}--\eqref{eq:matrix-twist} 
define the controlled topological extension.

The relation to hadron physics is motivational rather than identificational.  
The model shows how genuine three-body information can survive the removal of an explicit potential and enter the mass operator 
of a composite system.  It does not imply that $\theta_3$ is a QCD observable or that a one-dimensional contact term 
reproduces baryon spectroscopy.  Its value is to isolate a mechanism that can guide more realistic relativistic calculations.

A nonzero $\theta_3$ is also not, by itself, a physical CP-violating phase.  
A physical CP phase must survive all allowed field redefinitions and must be tested together with the transformed domain 
and discrete-symmetry operator.  A possible extension is to introduce flavor-space mass and three-body holonomy matrices.  
If they cannot be simultaneously made real and diagonal, rephasing-invariant relative phases may remain and generate 
CP-sensitive mixing in the composite mass matrix.  Such a flavor extension lies beyond the single-flavor construction studied here.

\section{Conclusion}
We formulated a relativistic three-body Dirac framework with Sakamoto-type pair interactions and an additional 
triple-coincidence structure.  The Poincar\'e algebra was stated as an operator constraint and reduced 
to explicit equations for the interaction-dependent boost.  The pairwise contact terms can be transferred 
to phase and matching data, while the unitary transformation rotates the matrix form of the nonzero constituent-mass operator.

The main result is the $Q_3$-sector mechanism.  The kinetic and pair-interaction parts commute 
with $Q_3=\alpha_1\alpha_2\alpha_3$, but the constituent-mass operator anticommutes with it.  
The matrix-valued three-body holonomy therefore assigns the opposite phases $e^{\pm i\theta_3}$ 
to two sectors that are mixed only when the constituent masses are nonzero.  
The block Hamiltonian and the corresponding two-state reduction display 
how this relative phase can enter a physical bound-state eigenvalue.

The six-sector loop construction translates this mechanism into the general compatibility condition
 $\det[W_3^{-1}\mathcal T_{\rm loop}-\mathbf1]=0$.  This is not yet an evaluated spectrum: 
 the energy-dependent normalizable sector bases and the complete matching matrices remain 
 to be constructed.  Accordingly, the present result establishes the structure 
 by which a three-body phase may modify the invariant mass, rather than a closed formula for that modification.

Finally, a bare multiplicative potential $g_3Q_3\delta(r)\delta(s)$ is not automatically equivalent 
to the topological connection whose curvature is supported at the origin.  
Establishing that equivalence requires a regulator, a self-adjoint limit, 
and a compatible boost generator.  The next concrete calculation 
is therefore to construct the six-sector transfer matrices for a specified regularization, 
evaluate $D(E,\theta_3)$, determine the mass-mixing blocks, and test Eqs.~\eqref{eq:boost1}--\eqref{eq:boost2}.  
Only at that stage can a closed or numerical curve $M_3(\boldsymbol g,\theta_3)$ and a nonzero coefficient $C_2(\boldsymbol g)$ be claimed.

\appendix
\section{Alternative projector structure in the three-body internal space}
\label{app:mixed_projector}
In the main text, we have adopted
\begin{equation}
Q_3=\alpha_1\alpha_2\alpha_3
\end{equation}
as the generator associated with the genuine three-body internal
transformation.  This choice involves the internal degrees of freedom of
all three particles multiplicatively and cannot be written as a simple
sum of pairwise operators.

For completeness, we consider here an alternative symmetric operator,
\begin{equation}
\widetilde Q_3
=
3-\alpha_1\alpha_2-\alpha_2\alpha_3-\alpha_3\alpha_1 .
\label{eq:Q3tilde}
\end{equation}
Although $\widetilde Q_3$ is not irreducible with respect to pairwise
structures, it provides a particularly simple characterization of the
internal sectors of the three-body Dirac system.

Using
\begin{equation}
\alpha_i^2=1,
\qquad
[\alpha_i,\alpha_j]=0
\qquad (i\neq j),
\end{equation}
one readily finds
\begin{equation}
\widetilde Q_3^{\,2}=4\widetilde Q_3.
\end{equation}
It is therefore natural to introduce
\begin{equation}
P_{\rm mix}
\equiv
\frac{\widetilde Q_3}{4}
=
\frac{1}{4}
\left(
3-\alpha_1\alpha_2-\alpha_2\alpha_3-\alpha_3\alpha_1
\right),
\label{eq:Pmix}
\end{equation}
which satisfies
\begin{equation}
P_{\rm mix}^2=P_{\rm mix}.
\end{equation}
Thus, $P_{\rm mix}$ is a projection operator.

Let
\begin{equation}
\alpha_i |s_1s_2s_3\rangle
=
s_i |s_1s_2s_3\rangle,
\qquad
s_i=\pm1 .
\end{equation}
The eigenvalue of $\widetilde Q_3$ is then
\begin{equation}
\widetilde q_3
=
3-s_1s_2-s_2s_3-s_3s_1 .
\end{equation}
For the two fully aligned configurations,
\begin{equation}
(+++) ,
\qquad
(---),
\end{equation}
one has
\begin{equation}
\widetilde q_3=0,
\qquad
P_{\rm mix}=0.
\end{equation}
For the remaining six configurations,
\begin{equation}
(++-),\quad
(+-+),\quad
(-++),\quad
(--+),\quad
(-+-),\quad
(+--),
\end{equation}
one obtains
\begin{equation}
\widetilde q_3=4,
\qquad
P_{\rm mix}=1.
\end{equation}
Hence $P_{\rm mix}$ projects onto the six-dimensional mixed internal
sector,
\begin{equation}
{\cal H}
=
{\cal H}_{\rm aligned}
\oplus
{\cal H}_{\rm mixed},
\end{equation}
with
\begin{equation}
\dim{\cal H}_{\rm aligned}=2,
\qquad
\dim{\cal H}_{\rm mixed}=6.
\end{equation}
This decomposition classifies the eight-dimensional internal spinor space into
a two-dimensional aligned subspace and a six-dimensional mixed subspace.  The
latter should not be confused with the six particle-ordering sectors of the
relative configuration space used in the matching construction; the two notions
of ``sector'' refer to different spaces.

A unitary transformation generated by this projector can be written as
\begin{equation}
U_{\rm mix}(\theta)
=
\exp\left(i\theta P_{\rm mix}\right).
\end{equation}
Since $P_{\rm mix}$ is a projector,
\begin{equation}
U_{\rm mix}(\theta)
=
(1-P_{\rm mix})
+
e^{i\theta}P_{\rm mix}.
\label{eq:Umix}
\end{equation}
Thus, the aligned sector remains unchanged, whereas the mixed
six-dimensional sector acquires the common phase $e^{i\theta}$.

The operator $P_{\rm mix}$ commutes with both the total momentum
\begin{equation}
P=\sum_{i=1}^{3}p_i
\end{equation}
and the kinetic part of the free Hamiltonian,
\begin{equation}
T=\sum_{i=1}^{3}\alpha_i p_i,
\end{equation}
namely,
\begin{equation}
[P_{\rm mix},P]=0,
\qquad
[P_{\rm mix},T]=0.
\label{eq:Pmix_kinetic}
\end{equation}
On the other hand, for the mass operator
\begin{equation}
M=m\sum_{i=1}^{3}\beta_i,
\end{equation}
one generally has
\begin{equation}
[P_{\rm mix},M]\neq0.
\label{eq:Pmix_mass}
\end{equation}
Consequently, the transformation (\ref{eq:Umix}) modifies the coupling
between the aligned and mixed internal sectors through the mass term.

Introducing
\begin{equation}
R_{\rm mix}=1-P_{\rm mix},
\end{equation}
the transformed mass operator takes the form
\begin{align}
U_{\rm mix}^{-1} M U_{\rm mix}
={}&
R_{\rm mix}MR_{\rm mix}
+
P_{\rm mix}MP_{\rm mix}
\nonumber\\
&+
e^{i\theta}
R_{\rm mix}MP_{\rm mix}
+
e^{-i\theta}
P_{\rm mix}MR_{\rm mix}.
\label{eq:mass_projector_transform}
\end{align}
Equation (\ref{eq:mass_projector_transform}) makes explicit that the
phase appears in the off-diagonal mass couplings between the two
internal sectors.

It is important, however, to distinguish this projector structure from
the genuine three-body operator used in the main text.  Indeed,
Eq.~(\ref{eq:Q3tilde}) can be rewritten as
\begin{equation}
\widetilde Q_3
=
(1-\alpha_1\alpha_2)
+
(1-\alpha_2\alpha_3)
+
(1-\alpha_3\alpha_1).
\label{eq:Q3tilde_pairwise}
\end{equation}
Thus, its internal Dirac structure is explicitly reducible to a sum of
pairwise operators.  For this reason, $\widetilde Q_3$ is not adopted
as the generator of the genuine three-body interaction in the main
analysis.

Nevertheless, $P_{\rm mix}$ is useful as an internal sector-classification
operator.  In particular, it separates the two fully aligned spinor states
from the six mixed spinor states.  This internal decomposition may be used
alongside the configuration-space matching construction, but it does not
project onto, or label, the six particle-ordering sectors themselves.

Finally, if $U_{\rm mix}$ is regarded only as a globally defined
unitary change of basis, the energy spectrum is unchanged by the
transformation.  A nontrivial phase dependence of the bound-state
energy requires that the phase be associated with a nontrivial global
matching condition or configuration-space holonomy around the
triple-coincidence point.  Thus, the role of $P_{\rm mix}$ in the
present work is primarily to clarify the sector structure, whereas the
irreducible operator
\begin{equation}
Q_3=\alpha_1\alpha_2\alpha_3
\end{equation}
remains the natural choice for representing the genuine three-body
internal structure discussed in the main text.

\section{Extension to $N$ particles}
For \(N\) distinguishable particles, two different generalizations of the three-body construction 
should be distinguished.

First, removing the triple-coincidence loci $x_i=x_j=x_k$ from configuration space gives 
the topological structure associated with three-particle exchanges. The corresponding 
fundamental group is the pure twin group \(PT_N\)\cite{HarshmanKnapp,Ohya2024}. Abelian one-dimensional representations may assign phases to generators
associated with triple-coincidence topology.  A simple flux construction can
introduce one parameter for each triple label, although for $N>3$ this does not
exhaust the most general one-dimensional representation of $PT_N$. A formal realization is

\begin{align}
\Phi\longmapsto
\exp\!\left[
i\sum_{i<j<k}
\theta_{ijk}Q_{ijk}\frac{\phi_{ijk}}{2\pi}
\right]\Phi,
\qquad
Q_{ijk}=
\alpha_i\alpha_j\alpha_k .
\end{align}

The number of triple labels in this simple construction is \(\binom{N}{3}\). 
As in the three-body problem, these operators commute 
with the kinetic matrices and anticommute with the corresponding constituent-mass matrices, 
providing a possible mechanism by which triple-sector phases become spectrally visible 
through mass-induced mixing.

A second, algebraically natural extension is a genuine \(N\)-body operator,
\begin{align}
Q_N=\prod_{i=1}^{N}\alpha_i ,
\qquad
Q_N^2=1 .
\end{align}
Since matrices belonging to different particles commute,
\begin{align}
[Q_N,\alpha_i]=0,
\qquad
\{Q_N,\beta_i\}=0 ,
\end{align}
and hence
\begin{align}
[Q_N,H_{\rm kin}]=0,
\qquad
\{Q_N,H_m\}=0 .
\end{align}
One may therefore formally introduce
\beq
W_N(\theta_N)=e^{i\theta_NQ_N},
\eeq
which assigns opposite phases $e^{\pm i\theta_N} $ to the $Q_N=\pm1 $ sectors, 
while nonzero constituent masses mix the two sectors. 
Thus the phase-sector-mixing mechanism identified for three particles possesses 
a direct algebraic \(N\)-body extension.

The topological interpretation, however, is special to the three-body case. 
After removal of the center-of-mass coordinate, the complete \(N\)-particle coincidence
 \(x_1=\cdots=x_N\) is a point in an \((N-1)\)-dimensional relative configuration space. 
 For \(N=3\), removal of this point gives \(\mathbb R^2\setminus\{0\}\) and 
 hence a nontrivial winding number. For \(N\ge4\), 
 by contrast, $\mathbb R^{N-1}\setminus\{0\}$ is simply connected, 
 so the full \(N\)-body coincidence does not by itself generate an Aharonov--Bohm-type 
 one-dimensional holonomy. The operators \(Q_{ijk}\) associated 
 with codimension-two triple-coincidence loci therefore provide 
 the natural topological extension, whereas \(Q_N\) should be regarded primarily 
 as an algebraic candidate for a genuine collective \(N\)-body phase.

As in the three-body case, any such extension must ultimately be supplemented 
by a self-adjoint domain and by an interaction-dependent boost that closes the Poincaré algebra.

\end{document}